\documentclass[twocolumn,showpacs,aps,pre,amsmath,amssymb,tightenlines,10pt]{revtex4}
\usepackage{graphicx}
\usepackage{dcolumn}
\usepackage{bm}

\def\be{\begin{equation}}
\def\ee{\end{equation}}
\def\bea{\begin{eqnarray}}
\def\eea{\end{eqnarray}}
\def\ba{\begin{array}}
\def\ea{\end{array}}

\def\0{$\sigma_0$}

\def\p{\phi}

\def\t{\theta}

\begin{document}

\title{A new mechanism for negative refraction and focusing using selective diffraction from surface corrugation}
\author{W. T. Lu}
\email{w.lu@neu.edu}
\author{Y. J. Huang, P. Vodo, R. K. Banyal, C. H. Perry}
\author{S. Sridhar}
\email{s.sridhar@neu.edu}
\affiliation{Department of Physics and Electronic Materials Research Institute,
Northeastern University, Boston, Massachusetts 02115}
\date{\today}

\begin{abstract}
Refraction at a smooth interface is accompanied by momentum transfer normal to the interface. We show that corrugating an initially smooth, totally reflecting, non-metallic interface provides a momentum kick parallel to the surface, which can be used to refract light negatively or positively. This new mechanism of negative refraction is demonstrated by visible light and microwave experiments on grisms (grating-prisms). Single-beam all-angle-negative-refraction is achieved by incorporating a surface grating on a flat multilayered material. This negative refraction mechanism is used to create a new optical device, a grating lens. A plano-concave grating lens is demonstrated to focus plane microwaves to a point image. These results show that customized surface engineering can be used to achieve negative refraction even though the bulk material has positive refractive index. The surface periodicity provides a tunable parameter to control beam propagation leading to novel optical and microwave devices.
\end{abstract}
\maketitle

\section{Introduction}

Negative refraction (NR) has been theoretically predicted \cite{Veselago} and experimentally realized \cite{Shelby,Parazzoli,Cubukcu,Parimi04,Parimi03}  in two types of materials.  One consists of simultaneously double negative permittivity and permeability material \cite{Pendry96,Pendry99,Smith04a,Shalaev}, leading to negative refractive index for the medium.  The other consists of a photonic crystal (PC) \cite{Joannopoulos95,Notomi,Luo02}, which is a periodic arrangement of scatterers in which the group and phase velocities can be in different directions leading to NR. In both cases, the bulk properties of the medium, which is inherently inhomogeneous, can be described as having an effective negative refractive index. These new types of materials lead to new applications such as perfect lens  \cite{Pendry00} and subwavelength imaging 
\cite{Luo03} .

In this paper we describe a new mechanism for achieving NR that utilizes surface corrugation at an initially smooth interface that is totally reflecting. Consider a plane wave incident on the corrugated surface of a material with bulk refractive index $n>1$  at an incident angle $\t$ (see Fig.1(a)). A single beam of light can be refracted negatively or positively by (i) making the incident angle greater than the critical angle  $\t>\t_c=\sin^{-1}(1/n)$ which suppresses the zeroth and all positive orders, and then (ii) tuning the corrugation wave vector to select the $-1$ order. We demonstrate this concept experimentally using grisms (grating prisms) at visible and microwave frequencies. We further generalize this concept to eliminate the critical angle restriction and achieve  all-angle-negative-refraction (AANR), which is demonstrated by negative lateral shift of an incident microwave beam by a flat multilayered structure with a surface grating. The presence of the surface grating changes the wave vector in the bulk medium and gives a new handle to control the emerging light from the interface. The corrugation provides a momentum kick to the incident light, enabling it to cross the interface and emerge refractively at angles that can be controlled. We further use this mechanism to create a new optical device: a transmission grating lens. As a prototype, we demonstrate focusing of plane waves by a plano-concave lens with a grating on the curved surface.

\section{Negative refraction with visible light and microwave}

\begin{figure}[htbp]
\center{\includegraphics [angle=0,width=8.5cm]{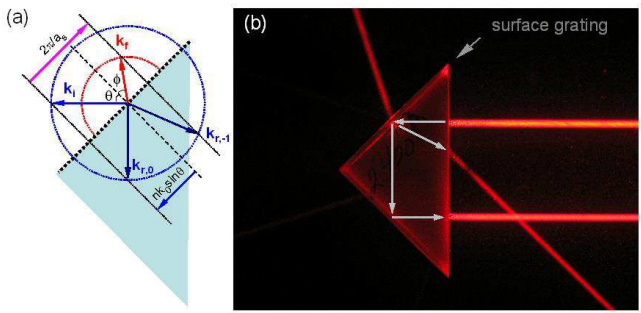}}
\caption{(a) Wave vector diagram for negative refraction from corrugated surface with grating period $a_s$. The red semicircle is the equi-frequency surface (EFS) in the air while the blue circle is the EFS in the dielectric. Here ${\vec k}_i$ is the incident wave vector in the glass,  ${\vec k}_f$ is the refracted one of minus one order in the air.  
 ${\vec k}_{r,0}$ and  ${\vec k}_{r,-1}$ are the reflected wave vectors in the dielectric of the zero-th and $-1$ order, respectively. 
(b) Optical experiment demonstrating negative refraction using a grism of size 2 cm with a grating density of 2400 lines/mm on the upper short surface. The He-Ne laser beam is normally incident to the hypotenuse. The solid lines with arrows indicate the propagation of the beams inside the grism.}
\label{fig1}
\end{figure}

The parallel component of the incident wave vector ${\vec k}$  along the surface is $k_{||}=nk_0\sin\t$. Here  $k_0=2\pi/\lambda$ is the wave number and $\lambda$ the wavelength in free space. Due to the surface corrugation of periodicity $a_s$, the wave vector along the grating surface is not conserved. The parallel components of the transmitted and reflected wave vectors along the grating surface are $k_{||m}=nk_0\sin\t+2m\pi/a_s$, according to the Floquet theorem \cite{Neviere}. Here $m$ is the order of the so-called Bragg waves. The radiating Bragg waves into the air have  
$-1<n\sin\t+m\lambda/a_s<1$. Otherwise, they will be evanescent, constituting the surface waves. For incident angles larger than the critical angle,  all the radiating Bragg waves have negative orders, $m<0$. Within the wavelength range 
$a_s(1+n\sin\t)/2<\lambda<a_s(1+n\sin\t)$, only the $m=-1$  Bragg wave will radiate from the grating surface into the air, which we call the refracted beam with wave vector ${\vec k}_f$  and $k_{f||}=nk_0\sin\t-2\pi/a_s$. For light within this range, an effective refractive index can be defined as 
\be
n_{\rm eff}=n-\lambda/(a_s\sin\t)
\ee
and Snell's law applies. If $a_s(1+n\sin\t)/2<\lambda<na_s\sin\t$, $n_{\rm eff}>0$ the refraction will be positive while for , $a_sn\sin\t<\lambda<a_s(1+n\sin\t)$, $n_{\rm eff}>0$ the refraction will be negative. This is illustrated in Fig. \ref{fig1}(a).

An experimental demonstration of NR at visible light using this mechanism is shown in Fig. 1(b). A holographic transmission grating with ruling density 2400 lines/mm and estimated groove depth $h \sim 130$ nm was replicated on one of the sides of an equilateral  right-angle BK7 prism of size 2 cm. A collimated laser beam is incident on the hypotenuse and passes through the grating surface of the grism. The incident angle at the grating is the prism angle $\t=\pi/4$  which is greater than the critical angle for the BK7 glass. Theoretical analysis indicates that only the $m=-1$  order beam would radiate into the air at negative angles if the incident light is within 435-860 nm wavelength range. Photographs of the experiments clearly indicate that the incident He-Ne (632.8 nm) laser beam ``refracts" negatively with angle $\p=27^\circ$ (see Fig. \ref{fig1}(b). Indeed this is indistinguishable from refraction by a prism made of a negative refractive index material with $n_{\rm eff}=-0.63$  for the red light. A sketch of the main beam trajectories inside the glass is shown in Fig. 1(b). All the beams can and have been explained by diffraction theory. NR was also observed on this grism for the green laser (532 nm) in which case 
$n_{\rm eff}=-0.29$.

The above experiments were repeated on an 1800 lines/mm grim with an estimated groove depth $h\sim 150$  nm. In the case of the red light with normal incidence to the hypotenuse, only the $m=-1$  beam will emerge negatively from the grating surface, whereas for the green light the  $m=-1$ order is diffracted positively with additional appearance of $m=-2$  order. Control experiments performed on regular prisms (without the grating) show positive refraction or complete reflection without any transmitted beam, depending on the incident angle, as is to be expected.

\begin{figure}[htbp]
\center{\includegraphics [angle=0,width=5.5cm]{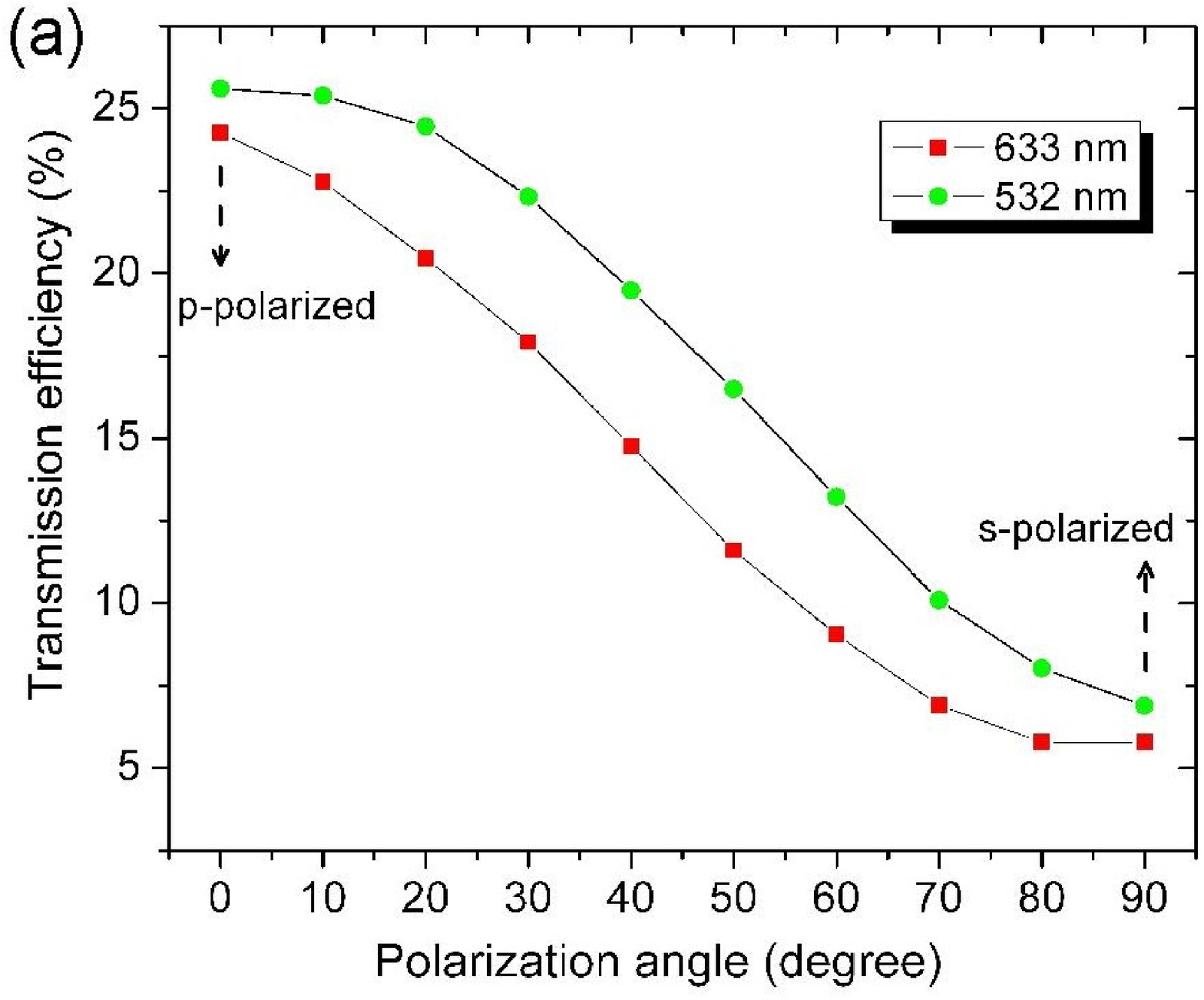}}
\center{\includegraphics [angle=0,width=6cm]{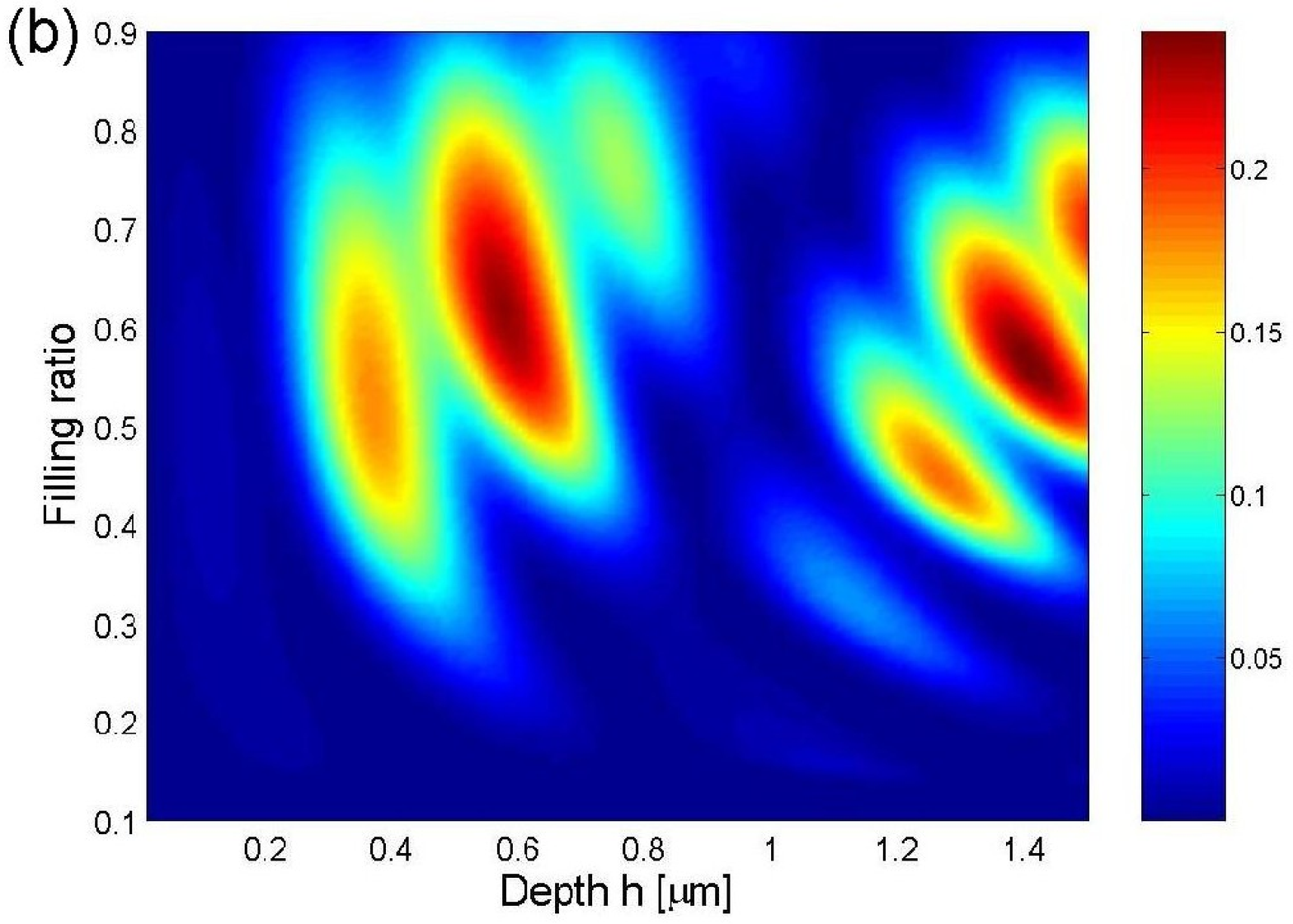}}
\center{\includegraphics [angle=0,width=6cm]{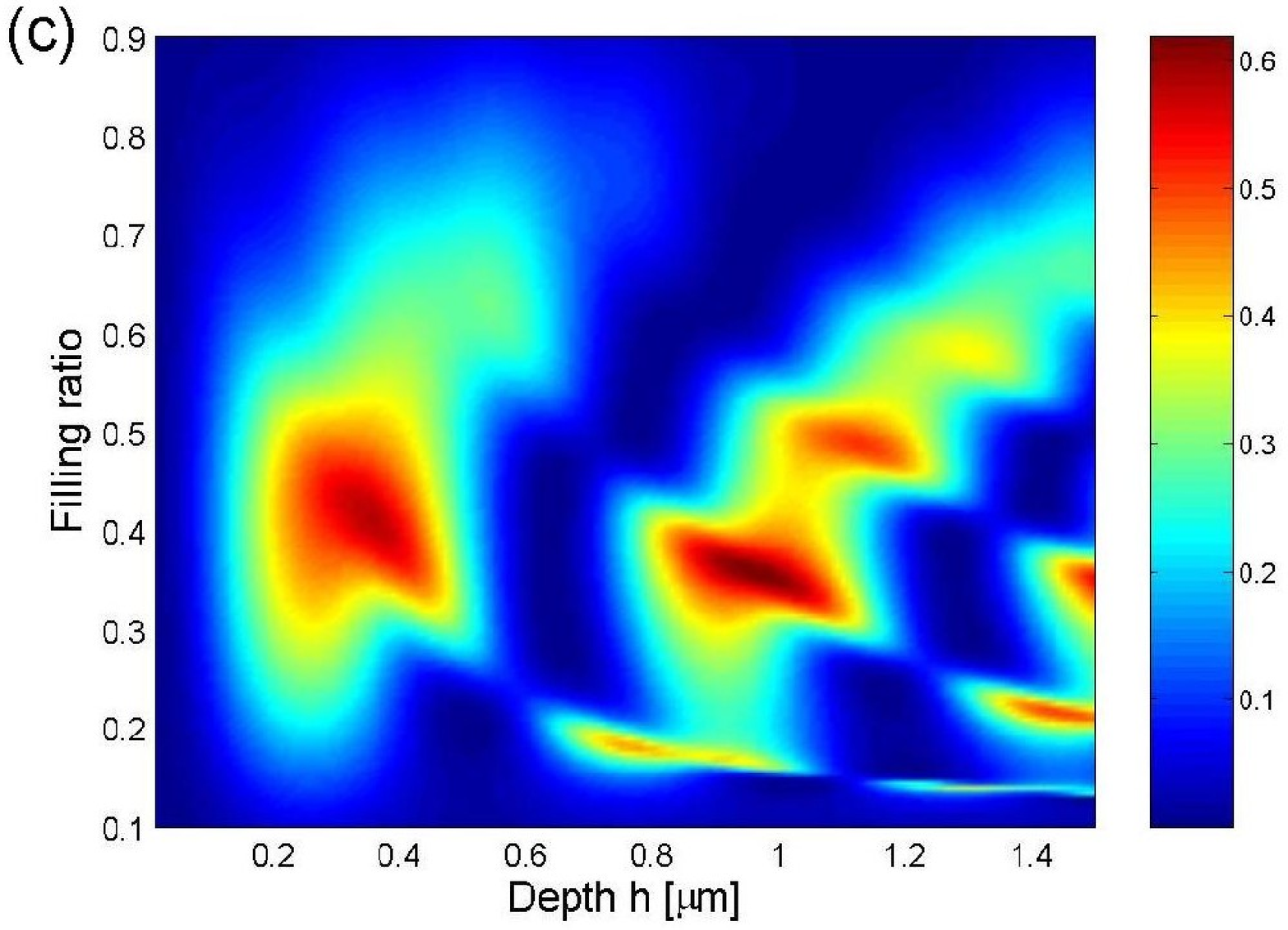}}
\caption{Transmission efficiency through a grism. (a) Experimental measurement of the transmission efficiency for different polarization of the BK7 grism with 2400 lines/mm. Calculated transmission efficiency of the S-polarization (b) and P-polarization (c) with incident angle $\pi/4$ for $\lambda=532$ nm through a lamellar grating on BK7 glass. The surface period of the lamellar grating has density 2400 lines/mm (as=416.7 nm).}
\label{fig2}
\end{figure}

For a fixed wavelength and groove geometry, the fraction of light diffracted into the  $m=-1$ order depends strongly on the polarization state of the incident light. The intensity transmission efficiency $\eta=I_{-1}/I_{in}$ for  $m=-1$ order was measured at different polarization orientations of the incident light.  A half-wave plate inserted between the grism and a polaroid is used to rotate the orientation of the linearly polarized light. For the $P$-polarization, the electric-field vector was parallel to the groove whereas for the $S$-polarization it was perpendicular to the grooves.  A maximum efficiency of 25\%  is attained for the 2400 lines/mm grism with the $P$-polarized green light.  The detailed transmission curves for different polarization orientations are shown in Fig. \ref{fig2}(a).

For practical applications, the efficiency of the power transmission is of primary concern. Since the direction of the refracted beam is determined only by the surface periodicity and not the detailed geometry of the grating, there is considerable freedom to design the grating surface to optimize the transmission. To exploit this freedom for transmission enhancement, we consider a specific grating, the lamellar grating. The transmission and reflection of waves were calculated using Bloch 
wave expansion method \cite{Lu07}. The transmission efficiency is plotted as a function of the groove depth h and the filling ratio for both the $S$- and $P$-polarizations at $\lambda=532$  nm as shown in Fig. \ref{fig2}(b) and \ref{fig2}(c), respectively. For groove depth  $h<50 \mu$m, the $P$-polarization can reach efficiency over 60\%, for $h\sim 1 \mu$m and filling ratio 0.34. For $h\sim 0.4 \mu$m and filling ratio is 0.4, the efficiency is 50\%.

\begin{figure}[htbp]
\center{\includegraphics [angle=0,width=5cm]{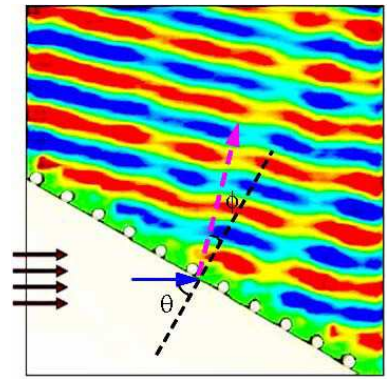}}
\caption{Microwave experiment demonstration of negative refraction by a polystyrene grism with surface grating period $a_s=2$  cm and incident angle $\t=\pi/3$ at 9 GHz. The solid arrows on the left indicate the incident direction of the microwave beam. The dashed line is the surface normal and the dashed arrow indicates the direction of propagation of the refracted beam.}
\label{fig3}
\end{figure}

The concept of NR by using selective diffraction is applicable over a wide range of frequencies. This is further exemplified by the grism experiments performed using microwaves as shown in Fig. \ref{fig3}. The grism consists of a right-angled polystyrene 
($\varepsilon=2.56$) prism with a surface grating of alumina rods next to the hypotenuse. The alumina rods have diameter 0.635 cm with grating periodicity  $a_s=2$ cm. The experiments were carried out in parallel-plate waveguide. The collimated microwave beam incidents normally to the shortest side of the prism and hits the hypotenuse with an incident angle $\t=\pi/3$. As shown in Fig. \ref{fig3} at 9 GHz, the beam emerges as if it was refracted negatively at an angle $\p=-16^\circ$, leading to $n_{\rm eff}=-0.32$. NR was observed between 6.3-10.8 GHz. Experimental data are in excellent agreement with theory and numerical simulations.

\section{Focusing by a plano-concave grating lens}
A unique feature of a negative index material is that it leads to focusing by a plano-concave lens \cite{Vodo05,Vodo06}. We show that the negative refraction mechanism demonstrated above can be used to design a plano-concave grating lens. For a plano-concave lens with circular curved surface of radius $R$, if the grating is placed such that the groove distance along the optical axis is a fixed number $a$, the surface periodicity will be $a_s=a/\sin\t$ . Here the angle $\t$ is the incident angle toward the curved surface. The effective refractive index is 
\be
n_{\rm eff}=n-\lambda/a
\ee
which is independent of $\t$. A focus is expected with a focal length $f(\t)=R[1+\sin\p/\sin(\t-\p)]$. The focal length depends on the angle $\t$, leading to aberration, which is present even in conventional lenses. The image quality is mainly impacted by a) the variation of the focal length for large incident angle $\t$ and b) the zero-th order diffraction which is present when $\t<\sin^{-1}(1/n)$. The strategy to improve the image quality is discussed next. 

\begin{figure}[htbp]
\center{\includegraphics [angle=0,width=8.5cm]{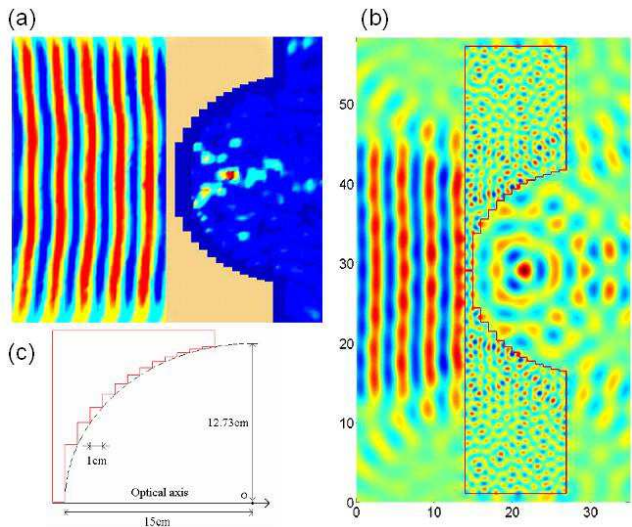}}
\caption{Demonstration of plano-concave grating lens focusing. (a) Composite figure of the microwave focusing experiment at 8.4 GHz using a plano-concave grating lens made of alumina with a grating on the curved surface. On the left the electric field of the incident beam measured without the presence of the grating lens is plotted. On the right the intensity of the electric field is plotted. In the middle is a photo of the lens. The grating lens behaves like a smooth plano-concave lens made of negative index material with  $n_{\rm eff}=-0.57$ at 8.4 GHz. (b) FDTD simulations at plano-concave lens without aberration made with $n=3$, $R = 15$ cm, and $a = 1$ cm at 8.5 GHz. Plotted is the electric field. The size of the system is in cm. (c) Details of the plano-concave lens (shown half). The dashed curve is an ellipse with semimajor 15 cm and semiminor 12.73 cm.}
\label{fig4}
\end{figure}

A good quality focus can be observed for the plano-concave grating lens with circular surface if $\lambda/a\sim n$,
$n_{\rm eff}\sim 0$, in which case the focal length $f(\t)$ is flat. For $|n_{\rm eff}|<1$, one can use a noncircular curve instead of a circular curve to minimize spherical aberration. This curve assumes an elliptical form  
$y^2/b^2+x^2=R^2$ with  $b=\sqrt{1-n_{\rm eff}^2}$ and $f=R(1+|n_{\rm eff}|)$  being the desired focal length. On this elliptical curve one places the grating such that the distance along the optical axis is a constant $a$ as in the circular case. 
In order to eliminate the diverging beam around the optical axis of the plano-concave lens due to the zero-th order diffraction, one can simply block this part of the lens. Even if the interference from the zero-th order diffraction could not be eliminated, it can be reduced.  For plano-concave lens with higher refractive index, this effect is smaller. For certain gratings on the plano-concave lens, the zero-th order diffraction can also be suppressed. For example for the staggered cut as in the one-dimensional (1D) PC , the part of the lens around the optical axis is flat as shown in the Fig. \ref{fig4}. For this grating one can choose the thickness of the lens around the optical axis such that the transmission through this part is a minimum. This is confirmed by numerical simulation.

An elliptical plano-concave grating lens made of alumina with semimajor axis $R=15$  cm and semininor axis 12.7 cm is shown in Fig. \ref{fig4}(c). The grating is made by staggered cuts on the concave surface, such that the horizontal distance of consecutive cuts is 1 cm. The effective index is $n_{\rm eff}=-0.53$ at 8.5 GHz, respectively. A high quality focus of a microwave beam is observed at 8.4 GHz as shown in Fig. \ref{fig4}(a). The inverse experiment was also performed in which a point source placed at the focal point will radiate a plane wave beam at 8.4 GHz. Numerical simulation (Fig. \ref{fig4}(b)) verifies both of the above mentioned focusing experiments, at 8.5 GHz. 

\section{All-angle-negative-refraction and negative lateral shift through grating multilayered structure}
So far we have demonstrated NR through the combination of total internal reflection (TIR) and selective negative diffraction by combining a surface grating with a homogeneous isotropic bulk material. The same NR mechanism introduced above is further applied in this section to achieve AANR \cite{Luo02} through a flat corrugated interface between air and an anisotropic medium. 

\begin{figure}[htbp]
\center{\includegraphics [angle=0,width=8cm]{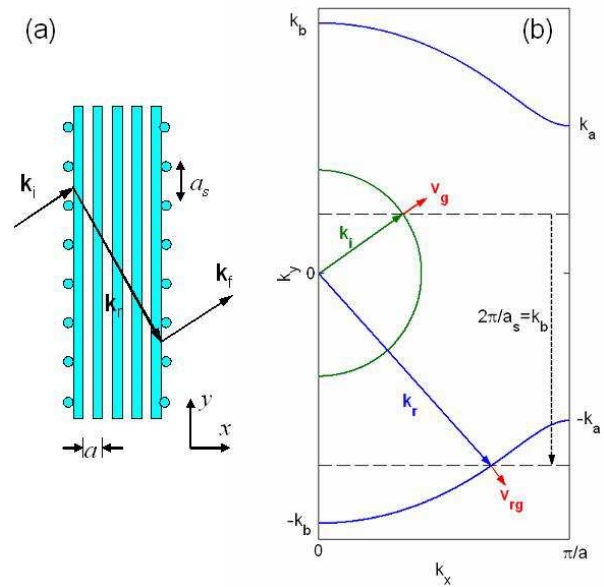}}
\caption{All-angle-negative-refraction using surface grating. (a) A slab of 1D multilayer PC of lattice spacing $a$, with surface grating $a_s$ on both surfaces. The surface grating gives rise to negative refraction for the 1D PC operating within the band gap. (b) Illustration of negative refraction using surface grating. The EFS of a 1D PC made of alumina bars with lattice spacing $a=0.9$ cm, 
bar thickness $d=0.5$ cm at 6.85 GHz is shown as the blue curves. The green semicircle is the EFS in the air.}
\label{fig5}
\end{figure}

To illustrate this principle to achieve AANR, we consider a multilayered structure which behaves as a one-dimensional photonic crystal (1D PC). For a 1D PC as shown in Fig. \ref{fig5}(a), there will be a band gap for normal incident plane waves within a certain frequency range. For these frequencies, transmission may be allowed for oblique incident angles. For example for the equi-frequency surface (EFS) of the 1D PC shown in Fig. \ref{fig5}(b), waves with incident angle $\t$, such as $k_a<k_0\sin\t<k_b$ will propagate. If for some frequencies, $k_0<k_a$ then for all the incident plane waves, there is TIR. So this 1D PC behaves as an omnidirectional mirror \cite{Fink}  for these E-polarized modes. If a grating with period as is introduced on the flat surface of the 1D PC,  for example with  $2\pi/a_s=k_a$, then a plane wave with an incident angle $\t$ will get a positive momentum kick along the surface. 
Thus the incident wave will couple to the Bloch wave with $k_y=k_0\sin\t+2\pi/a_s$  and propagate inside the 1D PC. 
However if $2\pi/a_s=k_b$, the incident wave will receive a negative momentum kick along the surface and couple to the Bloch wave with  $k_y=k_0\sin\t-2\pi/a_s$. And with proper design, it is possible that only the Bloch wave with $k_y=k_0\sin\t-2\pi/a_s$ will propagate inside the 1D PC. This refraction is well-defined and negative. Furthermore if $k_0=k_b-k_a$ , all the propagating waves will be transmitted into the 1D PC. Since $k_y$  of the Bloch wave is negative for every positive incident angle $\t$ this leads to a single-beam AANR. In this case, both the wave vector and group velocity refraction is negative. This scenario for NR is illustrated in Fig. \ref{fig5}(b). For 1D PCs different $k_y$  correspond to different modes. With the introduction of surface grating, Bloch states with $k_y$  and  $k_y+2m\pi/a_s$ are identical. Thus the 1D PC effectively becomes a 2D PC, resulting in a finite-sized first Brillouin zone of a rectangle shape. This is the simplest all-dielectric structure to achieve AANR. 

\begin{figure}[htbp]
\center{\includegraphics [angle=0,width=5.5cm]{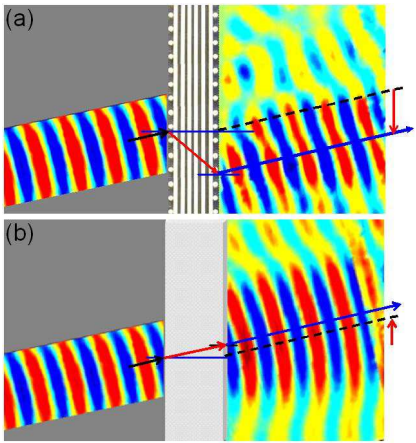}}
\caption{ (a) Experimental demonstration of negative lateral shift by a 1D PC with a surface grating, at 6.96 GHz. A 5.6 cm negative lateral shift was observed. The 1D PC is made of 6 layers of alumina bars with width $d=0.5$ cm and spacing  $a=1$  cm. The surface grating was formed by rods of the same material alumina with diameter 0.63 cm and spacing  $a_s=1.8$ cm. The width of the incident beam is 10 cm and the incident angle is $13.5^\circ$.  The incident and outgoing beams are plotted as the real part of the measured transmission coefficient $S_{21}$.
(b) Positive lateral shift for a microwave beam at 6.96 GHz by a slab of polystyrene with thickness 7.5 cm.}
\label{fig6}
\end{figure}

\begin{figure}[tbp]
\center{\includegraphics [angle=0,width=6cm]{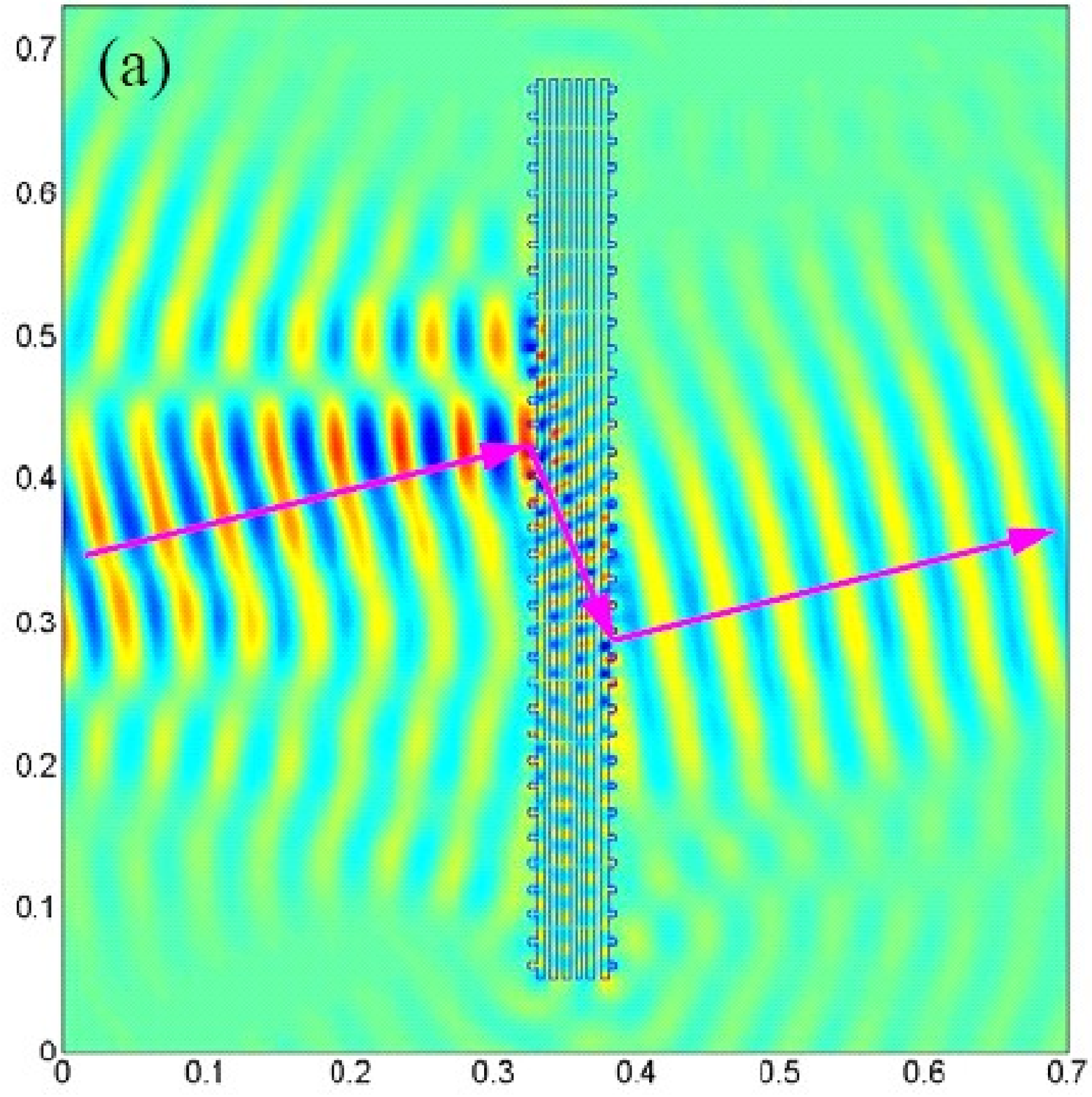}
\includegraphics [angle=0,width=6cm]{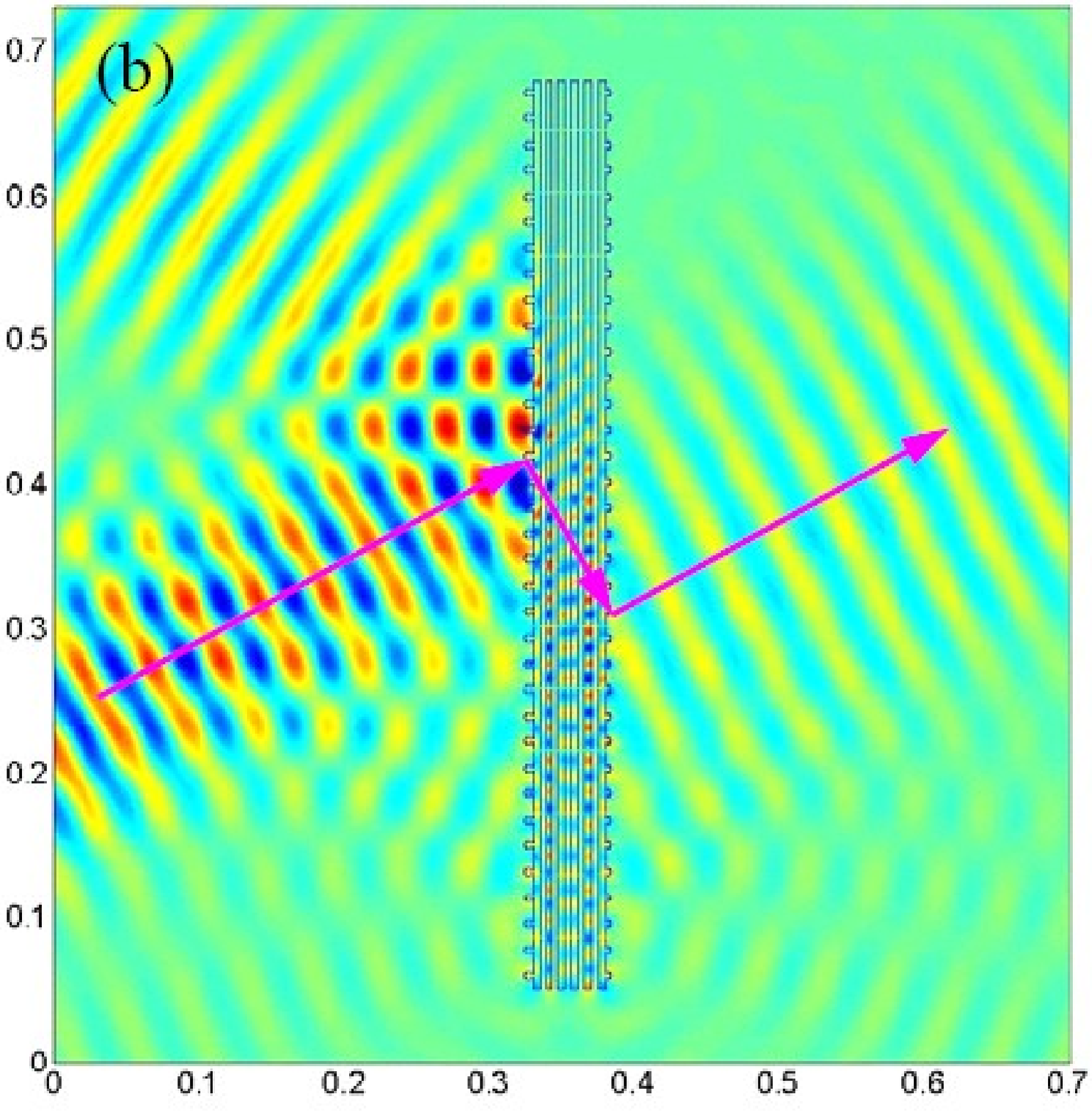}}
\caption{FDTD simulation of negative lateral shift of microwave beams through a grated 1D PC as specified in Fig. 6 at 6.96 GHz. 
(a) Microwave beam with incident angle $13.5^\circ$. (b) Microwave beam with incident angle $30^\circ$. The arrows indicate the energy flows of the incident and refracted beams. The size is in meters.}
\label{fig7}
\end{figure}

A microwave experiment carried out in parallel plate waveguide confirms the above mechanism for NR. Negative lateral shift was observed experimentally from 6.65-7.74 GHz for a grating multilayered structure. The 1D PC is made of 6 layers of alumina bars with thickness  $d=0.5$ cm, lattice spacing $a=0.9$  cm and surface grating $a_s=1.8$  (see Fig. \ref{fig6}). The incident angle of the 10 cm wide microwave beam was $13.5^\circ$. A 5.6 cm negative lateral shift is observed at 6.96 GHz, as shown in Fig. \ref{fig6}. Numerical simulations confirm AANR and negative lateral shift for a large range of incident angles for frequencies around 6.85 GHz as shown in Fig. \ref{fig7}.

\section{Discussion and conclusion}
This work provides a new perspective to the phenomenon of negative refraction. The NR can be seen as attributed to the bulk modification of the EFS and the surface grating. Surface periodicity alone is sufficient to achieve NR even with homogeneous positive index materials, as we have demonstrated. For 2D or higher dimension PC, the bulk periodic structure naturally introduces a surface periodicity at the interface. Improper surface modification of the PC may suppress or even diminish NR. Here we have shown that NR can be achieved by combining a surface grating with a multilayer 1D PC structure that is relatively easier to fabricate. Previous approaches to create NR materials with multilayer required the use of alternating layers of negative permittivity and negative permeability materials \cite{Fredkin,Alu} and have not yet been realized experimentally.  The realization of AANR in 1D PCs with surface corrugation opens a whole new realm of NR applications. Many structures such as photonic band gap materials currently used to guide waves or form cavities as photon insulators \cite{Yablonovitch,John,Lin,Ibanescu,Qi} can be modified to have NR and AANR through surface engineering. 

Although the phenomena presented in this paper are due to diffraction, ray optics does apply as we have shown in the design of the plano-concave grating lens (see Fig. \ref{fig4}). The mechanism of plano-concave lens focusing is different from that of the zone-plate where concentric rings are carved to give each ray the corrected phase and ray optics does not apply. Previous use of diffraction optics has been limited to reflection gratings. The grism used in astronomy  \cite{Loewen} is under the condition $n\sin\t<1$  thus allows the zero-th order diffraction. Our approach is different from the suppression of zero-th and enhancement of -1 order transmission through surface grating depth modification \cite{Noponen}. By removing the zero-th order Bragg diffraction completely, our work opens the door for new phenomena and applications such as plano-concave lens focusing and flat lens imaging. 

Surface engineering provides us a new dimension to manipulate waves. Refraction is an interfacial phenomenon, and this work shows that by controlled engineering of the interface, a totally reflecting surface can be made to refract negatively or positively, even though the materials utilized do not possess bulk negative refractive indices. The importance of surface modification has been previously recognized \cite{Decoopman,Smith04b} but has not been used as a mechanism to achieve negative refraction. 

To our knowledge, this work represents the first demonstration of negative refraction of visible light. The concepts discussed here are particularly suitable for integrated optical circuits, where the device dimension is about the size of the free space wavelength. The work reported here provides key ideas to harness diffraction to produce focusing devices. Thus new types of optical elements can be produced by using the above mechanism as a principle of design.

\section*{Acknowledgments}
Work supported by the Air Force Research Laboratory, Hanscom, MA through contract FA8718-06-C-0045 and the National Science Foundation through PHY-0457002.

\end{document}